\documentclass{article}
\usepackage[utf8]{inputenc}
\usepackage[a4paper]{geometry}
\usepackage{cite,graphicx,amsmath,amssymb}

\title{Coulomb Corrections for Bose--Einstein Correlations from One- And Three-Dimensional Lévy-Type Source Functions}

\author{Bálint Kurgyi$^1$, Dániel Kincses$^1$, Márton Nagy$^1$, Máté Csanád$^1$\\
E{\"o}tv{\"o}s Lor{\'a}nd University, Institute of Physics,\\
H-1117 Budapest, P{\'a}zm{\'a}ny P. s. 1/A, Hungary}

\begin{document}
\maketitle
\begin{abstract}
In the study of femtoscopic correlations in high-energy physics, besides Bose--Einstein correlations, one has to take final-state interactions into account. Amongst them, Coulomb interactions play a prominent role in the case of charged particles. Recent measurements have shown that in heavy-ion collisions, Bose--Einstein correlations can be best described by L\'evy-type sources instead of the more common Gaussian assumption. Furthermore, three-dimensional measurements have indicated that, depending on the choice of frame, a deviation from spherical symmetry observed under the assumption of Gaussian source functions persists in the case of L\'evy-type sources. To clarify such three-dimensional L\'evy-type correlation measurements, it is thus important to study the effect of Coulomb interactions in the case of non-spherical L\'evy sources. We calculated the Coulomb correction factor numerically in the case of such a source function for assorted kinematic domains and parameter values using the Metropolis--Hastings algorithm and compared our results with previous methods to treat Coulomb interactions in the presence of L\'evy sources.
\end{abstract}
\thispagestyle{empty}

\section{Introduction}
The investigation of Bose--Einstein or HBT correlations offers a way to gain information about the space-time dynamics of heavy-ion collisions on the femtometer scale. Such information can lead to a better understanding of the space-time geometry of the collision and particle production mechanisms and could even indicate critical phenomena~\cite{Csorgo:1999sj,Bolz:1992hc,Adamczyk:2014mxp,Adare:2017vig}.

For the study of Bose--Einstein correlation functions, one usually makes an assumption for the source function. There is now a large amount of evidence showing that in heavy-ion collisions, there is indeed a significant deviation from Gaussian shape, such as source imaging results showing a long-range, power-law-type component~\cite{Adler:2006as,PHENIX:2007grx,Shapoval:2013bga}. It turns out that a suitable choice is a L\'evy-type source function~\cite{Adare:2017vig,NA61SHINE:2023qzr,CMS:2023xyd}.

In one-dimensional correlation measurements, the correlation function is measured as a function of only one relative momentum variable, the magnitude of the momentum difference.
This type of measurement assumes the spherical symmetry of the spatial source (and thus the momentum space correlation) and is more suitable for situations wherein a lack of experimental statistics prevents the detailed mapping of the momentum space. Three-dimensional measurements, on the other hand, can yield further information about the space-time geometry of the source, so whenever experimental statistics make it possible, it is desirable to perform such measurements. Building on the substantial progress that three-dimensional Gaussian measurements have made in the understanding of the space-time structure of particle production in heavy-ion collisions, it is also of interest to perform L\'evy-type measurements in a three-dimensional setting. The first such measurement has already been reported~\cite{Kurgyis:2018zck}.

In the present paper, we aimed at developing a methodology for such a measurement.
Coulomb corrections are an essential ingredient of all HBT correlation measurements that use identical charged particles, as most do: the final-state Coulomb repulsion of the outgoing particles modifies the shape of the observed correlation function in a complicated manner, and in experimental analyses, one usually applies a correction factor, the Coulomb correction, to account for this effect~\cite{Sinyukov:1998fc,Bowler:1991vx}. At present, the Coulomb correction for L\'evy distributions is available only in the spherically symmetric case~\cite{Csanad:2019lkp}. Our goal was two-fold: first, we investigated the Coulomb correction for three-dimensional L\'evy sources and determined a sound method for its use in experimental work. Second, in doing so we encountered the question of the proper choice of coordinate frame, namely the longitudinally co-moving system (LCMS) and pair center of mass system (PCMS) of the particle pair. We thus investigated the implications of using these coordinate frames for the measurements and calculations.

\subsection{Two-Particle Correlation Functions}

The $n$-particle correlation functions are defined as

\begin{equation}
C_n = \frac{N_n(k_1,\cdots k_n)}{\prod_{i=1}^nN_1(k_i)},    
\end{equation}
where $N_n$ is the $n$-particle invariant momentum distribution. In a statistical picture, one introduces the source function $S(x,k)$ that characterizes the particle production at a given space-time point $x$ and momentum $k$, and writes up the $N_n(k_1,\dots,k_n)$ distribution using this function and the $n$-particle wave function $\psi_n(x_1,\cdots x_n,k_1,\cdots k_n)$ as

\begin{equation}
N_n(k_1,\cdots k_n) = \int |\psi_n(x_1,\cdots x_n,k_1,\cdots k_n)|^2
\prod_{i=1}^n S(x_i,k_i)\mathrm{d} x_i.
\end{equation}
In particular, for single-particle distributions, we have $|\psi_1|^2=1$. Thus,

\begin{equation}
N_1(k) = \int\mathrm{d}x\,S(x,k),    
\end{equation}
and so one obtains the two-particle correlation function as

\begin{equation}
C_2(k_1,k_2) = \frac{\int\mathrm{d}x_1\mathrm{d}x_2\,S(x_1,k_1)S(x_2,k_2)|\psi_2(x_1,k_1,x_2,k_2)|^2}{\int\mathrm{d}x_1\,S(x_1,k_1)\,\int\mathrm{d}x_2\,S(x_2,k_2)}.
\end{equation}

We introduced the average and relative space-time and momentum variables as
\begin{equation}
q \equiv k_1-k_2,\qquad K\equiv\frac{k_1+k_2}2,\qquad \rho\equiv x_1-x_2,\qquad R = \frac{x_1+x_2}2.
\end{equation}

Although the two-particle wave function appearing in the above formula itself is a function of all relevant variables, its modulus square depends only on $\rho$ and $q$, as shown below. We assumed this in advance and used a notation that reflected this by suppressing the $K$ and $R$ variables in the notation of $|\psi_2|^2$.

The overall normalization of the $S(x,k)$ source function cancels from the $C_2$ correlation function, so from now on we treat it as unity, i.e., we write
\begin{equation}
C_2(q,K) = \int|\psi_2(q,\rho)|^2 S(R-{\textstyle\frac12}\rho,K-{\textstyle\frac12}q)S(R+{\textstyle\frac12}\rho,K+{\textstyle\frac12}q)\,\mathrm{d}\rho\,\mathrm{d}R.
\end{equation}

A usual approximation is that one neglects the $\pm\frac12k$ in the arguments of the source functions; in other words, one approximates the $k_1$ and $k_2$ momenta in the source functions as $k_1\approx k_2\approx K$. In doing so, we noted that it was useful to introduce the relative coordinate distribution, also called pair distribution, $D(\rho,K)$, which is the auto-convolution of the source function in the first variable:
\begin{equation}
D(\rho,K) = \int \mathrm{d}R\,S\big(R+{\textstyle\frac12}\rho,K\big)
S\big(R-{\textstyle\frac12}\rho,K\big),
\end{equation}
with which the two-particle correlation function could then be expressed as
\begin{equation}
C_2(q,K) = \int  |\psi_2 (q,\rho)|^2 D(\rho,K) \mathrm{d} \rho.
\end{equation}

\subsection{L\'evy Sources}

For the source function, we assumed a symmetric L\'evy distribution~\cite{Csorgo:2003uv}:
\begin{align}
    S(r,K)=\mathcal{L}^{(4D)}(r^\mu,\alpha(K),R^2_{\sigma\nu}(K))
    =\int \frac{\mathrm{d}^4 q}{(2\pi)^4} e^{iq_\mu r^\mu} e^{-\frac{1}{2}|q^\sigma 
    R^2_{\sigma\nu} q^\nu|^{\alpha/2}},
\end{align}
where $\alpha$ is the L\'evy exponent, and $R^2_{\sigma\nu}$ is a two-index symmetric tensor containing the squares of the L\'evy scale parameters. The momentum dependence of the source was assumed to manifest itself through the momentum dependence of these parameters. For such a L\'evy-type source, the relative coordinate distribution $D(r,K)$ is itself a L\'evy distribution with the same $\alpha$ but with scale parameters modified as $R^2 \rightarrow 2^{2/\alpha} R^2$.

By choosing a reference frame and making some assumptions, we constrained the form of the $R^2_{\sigma\nu}$ matrix. In the case of Bose--Einstein correlation measurements in high-energy heavy-ion collisions, one sets the laboratory frame as the center-of-mass frame of the colliding nuclei. Most two-particle Bose--Einstein measurements are carried out with respect to the so-called longitudinally co-moving system (LCMS) (see, e.g., Refs.~\cite{Adamczyk:2014mxp,Adare:2017vig}), which is defined as the frame that is connected to the laboratory frame by a Lorentz boost along the collision axis ($z$ axis), with the criterion that the longitudinal component of the average momentum of the particle pair $K_z$ vanishes in this frame.

We made the assumption that our source could be described by a spatially three-dimensional symmetric L\'evy shape with only diagonal terms in the scale parameter matrix $R^2_{\sigma\nu}$, and that the freeze-out was simultaneous in the LCMS frame. We saw that the momentum variable of the source function translated essentially as the average momentum of the particle pair (after the $k_1\approx k_2\approx K$ approximation), so for $S(x,K)$, we determined the mean LCMS frame as the frame wherein its $K$ variable has no $K_z$ component. Therefore, the $R^2_{\sigma\nu}$ tensor has the following form:
\begin{align}
    R^2_{\sigma\nu}=
    \begin{pmatrix}
    0&0&0&0\\
    0&R_\text{out}^2&0&0\\
    0&0&R_\text{side}^2&0\\
    0&0&0&R_\text{long}^2
    \end{pmatrix},
\end{align}
where out, side, and long indicate that we used Bertsch--Pratt coordinates~\cite{Pratt:1990zq,Bertsch:1988db}. We could then simplify the four-dimensional 
L\'evy distribution as a product of a Dirac delta function and a three-dimensional symmetric L\'evy distribution:
\begin{align}\label{eq:levy_lcms}
    \mathcal{L}^{(4D)}&=\delta (t^{L}) \mathcal{L}^{(3D)}(\Vec{r}^L,\alpha,R_\text{out},
    R_\text{side},R_\text{long}),\\
    \mathcal{L}^{(3D)}(\Vec{r}^L,\alpha,R_\text{out},R_\text{side},R_\text{long})&=
    \int \frac{\mathrm{d}^3q}{(2\pi)^3}e^{-i\vec{q}\vec{r}^L}e^{-\frac12|q_\text{out}^2
    R_\text{out}^2+q_\text{side}^2R_\text{side}^2+q_\text{long}^2R_\text{long}^2|^{\alpha/2}},
\end{align}
where the $L$ superscript indicates that these coordinates are in the LCMS. As noted above, for such a source function, the pair distribution is a L\'evy distribution with modified scale parameters:
\begin{align}\label{eq:Dlevy_lcms}
    D(\vec{r}^L,K) &=\delta (t^{L}) \mathcal{L}^{(3D)}(\Vec{r}^L,\alpha,2^{\frac1\alpha}R_\text{out},
    2^{\frac1\alpha}R_\text{side},2^{\frac1\alpha}R_\text{long}),
\end{align}

From this, in the case when final-state interactions were neglected and thus the final-state wave function was a symmetrized plane wave, we could easily obtain the form of the two-particle correlation function in the LCMS with the above-mentioned source by means of an (inverse) Fourier transform \cite{Csorgo:2003uv}:
\begin{align}
    C_2^{(0)}(\vec{q},\alpha,R_\text{out},R_\text{side},R_\text{long})=1+
    e^{-|q_\text{out}^2R_\text{out}^2+q_\text{side}^2R_\text{side}^2+
    q_\text{long}^2R_\text{long}^2|^{\alpha/2}}.
\end{align}

\section{Methodology}

\subsection{Coulomb Interaction}

To take into account the Coulomb interaction, one has to use the Coulomb interacting two-particle wave function. This is the solution of the two-particle Schrödinger equation with a repulsive Coulomb force and the appropriate boundary conditions at infinity.

Utilizing the Schrödinger equation implies a non-relativistic treatment, which is a justifiable approximation in the PCMS (pair co-moving system) frame, i.e., the center-of-mass frame of the two particles. The solution of the Schrödinger equation of interest to us is written as~\cite{Sinyukov:1998fc,Landau3:1977}
\begin{align}
\begin{split}
    \psi(\vec{R}^P,\vec{r}^P,\vec{K}^P,\vec{k}^P)=\frac{N}{\sqrt{2}}e^{-i2\vec{K}\vec{R}}
    \big[e^{i\vec{k}\vec{r}}F(-i\eta,1,i(kr-\vec{k}\vec{r}))+\\
    +e^{-i\vec{k}\vec{r}}F(-i\eta,1,i(kr+\vec{k}\vec{r}))\big],
\end{split}
\end{align}
where a symmetrization has been performed, as required for pairs of identical bosons.
In this expression, $F(a,b,z)$ is the confluent hypergeometric function, $\vec{k}=\vec{q}/2$, $k=|\vec{k}|$, and
\begin{align}
    \eta&=\frac{m c^2 \alpha}{2\hbar c k}, &
N&=e^{-\frac{\pi\eta}{2}}\Gamma(1+i\eta),
\end{align}
where $\alpha$ is the fine-structure constant; $m$ is the particle mass (e.g., pion mass); and $\Gamma(z)$ is the gamma function.

To evaluate the two-particle correlation function, we needed the modulus square of the wave function, with which the $\vec{R}$ and $\vec{K}$ dependence was lost (as mentioned earlier):
\begin{align}\label{eq:wave_func_norm}
\begin{split}
    |\psi(\vec{r}^P,\vec{k}^P)|^2=\frac{2\pi\eta}{e^{2\pi\eta}-1}\cdot\frac12\cdot\Big[|F(-i\eta,1,i(kr+
    \vec{k}\vec{r}))|^2+\\
    +e^{2i\vec{k}\vec{r}}F(-i\eta,1,i(kr-\vec{k}\vec{r}))F(i\eta,1,-i(kr+\vec{k}
    \vec{r}))\Big] + (\vec r\leftrightarrow-\vec r).
\end{split}.
\end{align}

To arrive at the two-particle correlation function, one has to evaluate a $\mathrm{d}^4r$ integral over the whole space-time. This can be performed in any coordinate frame, but our approximations detailed above made strong arguments in favor of some preferred coordinate systems. However, even with these in mind, we had several options to explore:
\begin{enumerate}
    \item We could assume that the $R^2_{\sigma\nu}$ matrix, and thus the whole source function, is the same in the PCMS and the LCMS frames. This is essentially an approximation where $\vec{K}\approx0$. However, this is a rather strong approximation, and one of the goals of HBT measurements is indeed to explore the average momentum (or transverse mass) dependence of the parameters that describe the source.
    \item There are two objects, one in the PCMS (the wave function) and the other in the LCMS (the source function). We could try to transform the wave function from the PCMS to the LCMS and then use the simple form of the source function and obtain the result in LCMS coordinates. However, the two-particle wave function of Equation~\eqref{eq:wave_func_norm} is not a relativistic expression; thus, we refrained from trying to come up with the right transformation of this object.
    \item The third option was to evaluate the integral in the PCMS, as the two-particle Coulomb wave function is only known in the PCMS. This meant that the L\'evy source had to be transformed from the LCMS to the PCMS.
\end{enumerate}
Below, we proceed with the third option listed above. We introduce some further notations: the average transverse momentum in the LCMS  $K_T$, the transverse mass $m_T=\sqrt{m^2+K_T^2}$, and the $\beta_T=K_T/m_T$ factor. The Lorentz boost from the LCMS to the PCMS is then

\begin{equation}
{\Lambda}_{\mu}^{\nu}=\frac{1}{m}
\begin{pmatrix}
m_T&-K_T&0&0\\
-K_T&m_T&0&0\\
0&0&m&0\\
0&0&0&m
\end{pmatrix}.
\end{equation}

The L\'evy distribution then transforms as a scalar from the LCMS to the PCMS, meaning we had to evaluate Equation (\ref{eq:levy_lcms}) at the coordinates $r'=\Lambda^{-1} r$, where the 
transformation is the following:
\begin{equation}
\begin{pmatrix}
t^{L}\\
\vec{r}^{L}
\end{pmatrix}=\frac{1}{m}
\begin{pmatrix}
m_T t^{P} +K_T r^{P}_\text{out}\\
K_T t^{P} +m_T r^{P}_\text{out}\\
m r^P_\text{side}\\
m r^P_\text{long}
\end{pmatrix}.
\end{equation}

The temporal integral could then be easily evaluated and, subsequently, knowing the form of $D(\vec r,K)$ from Equation~(\ref{eq:Dlevy_lcms}) above, we were left with the following expression  (where $2\vec{k} = \vec{q}$):
\begin{equation}
    C_2^{(C)}(\vec{q})=\int \mathrm{d}^3r |\psi(\vec{k},\vec{r})|^2\mathcal{L}^{(3D)}
    \left(\sqrt{1{-}\beta_T^2}r_\text{out},r_\text{side},r_\text{long},\alpha,2^{\frac1\alpha}R_\text{out},2^{\frac1\alpha}R_
    \text{side},2^{\frac1\alpha}R_\text{long}\right),
\end{equation}
where we dropped the $P$ superscripts for simplicity, but every momentum and spatial coordinate is in the PCMS. Furthermore, we could utilize a simple scaling relation of the three-dimensional L\'evy distribution:

\begin{equation}
\begin{split}
    \mathcal{L}^{(3D)}\left(\sqrt{1-\beta_T^2}r_\text{out},r_\text{side},r_\text{long},
    \alpha,2^{\frac1\alpha}R_\text{out},2^{\frac1\alpha}R_\text{side},2^{\frac1\alpha}R_\text{long}\right) \sim \\
    \sim\mathcal{L}^{(3D)}\left(\vec{r},\alpha,2^{\frac1\alpha}R_\text{out}/\sqrt{1-\beta_T^2},2^{\frac1\alpha}R_
    \text{side},2^{\frac1\alpha}R_\text{long}\right),
\end{split}
\end{equation}
a relation that could be derived by scaling the $\vec q$ integration variable in the definition of $\mathcal L^{(3D)}$, Equation~(\ref{eq:levy_lcms}). In this equation, $\sim$ stands for proportionality, and constant factors in $S(x,k)$ cancel from the two-particle correlation function. Thus, the integral we intended to calculate was

\begin{equation}\label{eq:final_int}
    C_2^{(C)}(\vec{q},\alpha,R_1,R_2,R_3)=\int \mathrm{d}^3r |\psi(\vec{k},\vec{r})|^2
    \mathcal{L}^{(3D)}\left(\vec{r},\alpha,R_1,R_2,R_3\right),
\end{equation}
where $R_1 = 2^{\frac1\alpha}R_\text{out}/\sqrt{1-\beta_T^2}$, $R_2=2^{\frac1\alpha}R_\text{side}$, $R_3=2^{\frac1\alpha}R_\text{long}$. 
This expression could be evaluated numerically.

\subsection{Numerical Simulations}

For the evaluation of the integral, we utilized the Metropolis--Hastings algorithm.
This algorithm can be used to evaluate integrals of the form 

\begin{equation}
    I=\int_\Omega \mathrm{d}x f(x)\cdot g(x),
\end{equation}
where $f(x)$ can be thought of as a probability distribution, and $g(x)$ is the function of interest~\cite{Metropolis:1953am,Hastings:1970aa}. In our case, the three-dimensional symmetric L\'evy distribution was the probability distribution, and the function of interest was from Equation~(\ref{eq:wave_func_norm}):
\begin{align}
    f(x) \mathrm{d}x &:= \mathcal{L}^{(3D)}\left(\vec{r},\alpha,R_1,R_2,R_3\right) 
    \mathrm{d}^3r,\\
    g(x) &:= |\psi(\vec{k},\vec{r})|^2.
\end{align}

We could utilize two transformations. First, with the reflection relations of the confluent hypergeometric functions, we used the second term in Equation~(\ref{eq:wave_func_norm}):
\begin{equation}
\begin{split}
    e^{2i\vec{k}\vec{r}}F(-i\eta,1,i(kr-\vec{k}\vec{r}))F(i\eta,1,-i(kr+\vec{k}\vec{r}))=\\
    =F(1+i\eta,1,-i(kr-\vec{k}\vec{r}))F(1-i\eta,1,-i(kr+\vec{k}\vec{r})).
\end{split}
\end{equation}

Additionally, we could transform the 3D symmetric L\'evy distribution:
\begin{align}
\mathcal{L}^{(3D)}(\vec{r},\alpha,R_1,R_2,R_3)&=\frac1{R_1R_2R_3}\mathcal{L}^{(1D)}(s(\vec{r}),
\alpha,1),
\end{align}
where
\begin{align}
s(\vec{r})&=\sqrt{\frac{r_\text{out}^2}{R_1^2}+\frac{r_\text{side}^2}{R_2^2}+\frac{r_
\text{long}^2}{R_3^2}},
\end{align}
and $\mathcal{L}^{(1D)}$ is the spherically symmetric version of the three-dimensional L\'evy distribution, as a function of the radial variable. Its its expression is then
\begin{align}
\mathcal{L}^{(1D)}(x,\alpha,1) = \frac1{2\pi^2x}\int_0^\infty dq\,q^2\sin(qx)e^{-\frac12q^\alpha}.
\end{align}

We could thus perform the integral in Equation~(\ref{eq:final_int}); this was carried out using spherical coordinates on the domain $\Omega=[0,r_\text{max}]\times[0,2\pi]\times[0,\pi]$, with an $r_\text{max}$ chosen so that the integral of the L\'evy  distribution ($I=\int \mathcal{L}$) was a maximum of $1\%$ less than $1$ ($I\geq0.99$).

\section{Results}\label{sec:results}

First, we compared our three-dimensional calculations and other available, spherically symmetric calculations for L\'evy sources. Then, we investigated the implications of the fact that most measurements are in the LCMS, and the source is assumed to be spherical there for one-dimensional analyses, but the integral of Equation~\eqref{eq:final_int} is in the PCMS.

\subsection{Three-Dimensional Calculations}

\begin{figure}
    \includegraphics[width=0.5\textwidth]{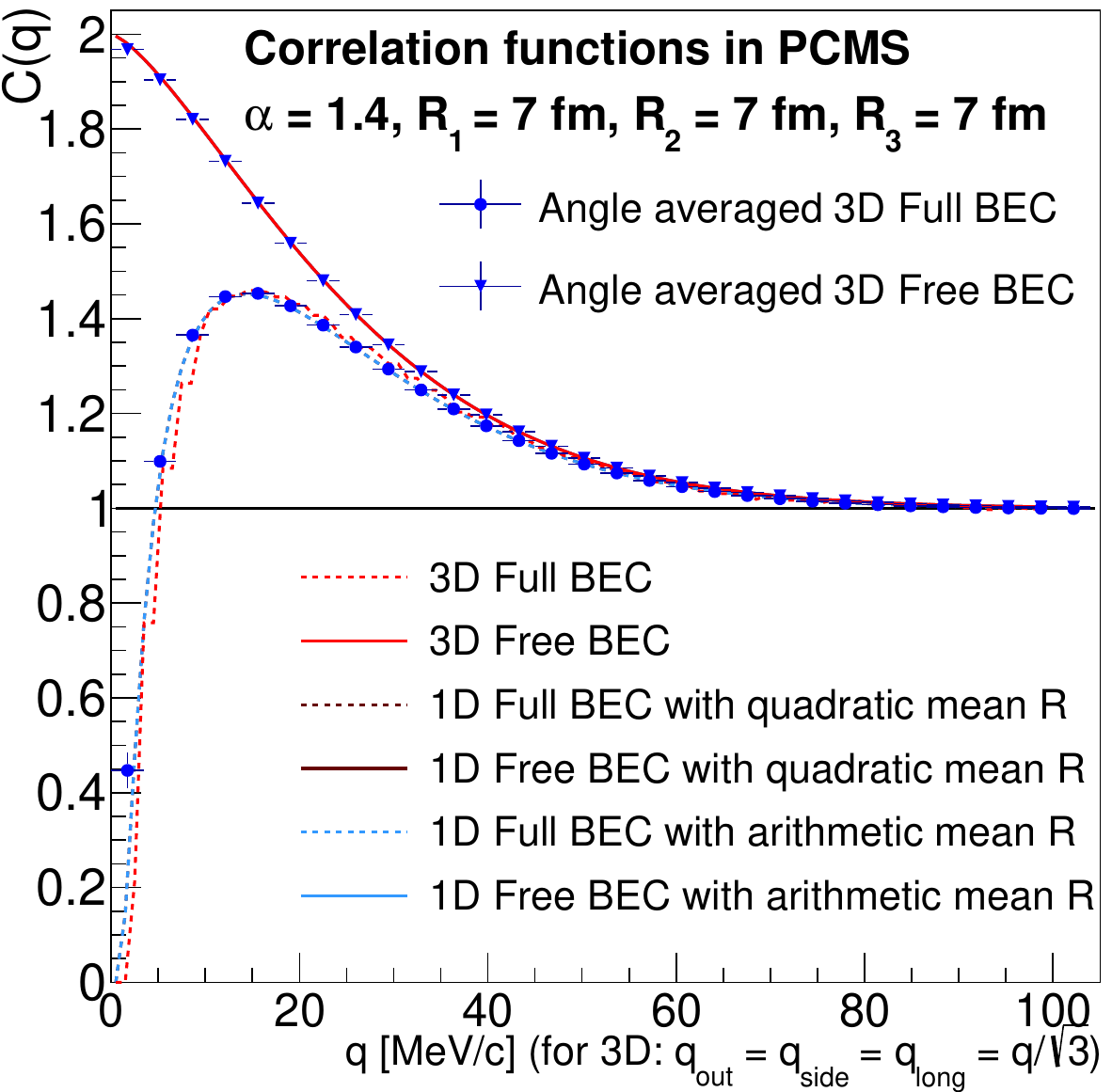}\hfill
    \includegraphics[width=0.5\textwidth]{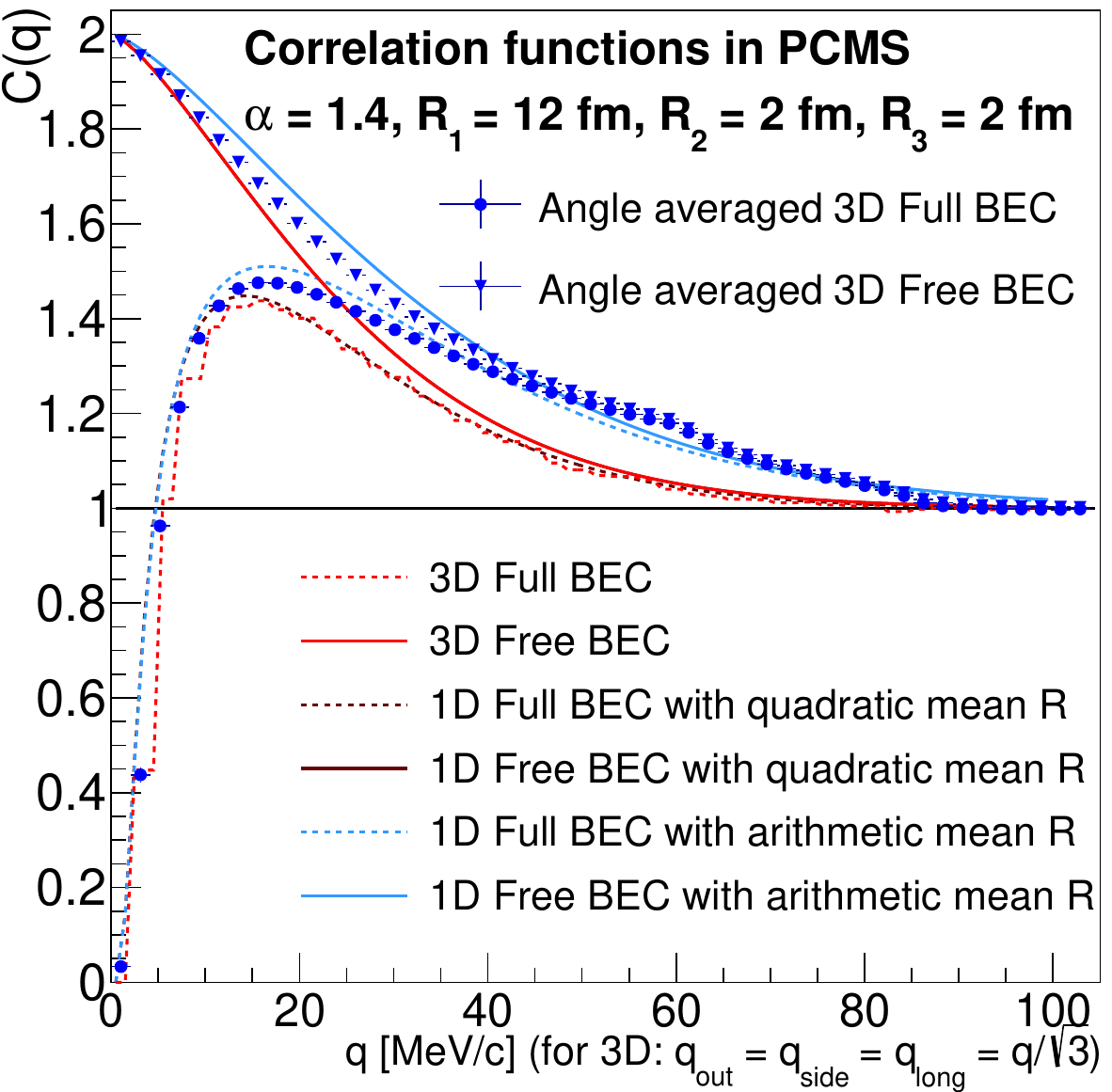}
    \caption{On the left-hand side, the two-particle correlation functions are shown in a spherical case for the three-dimensional calculation in comparison with one-dimensional calculations in the presence of Coulomb interactions in final-state interactions. On the right-hand side, a non-spherical three-dimensional calculation is shown alongside one-dimensional calculations with quadratic and arithmetic mean scale parameters.}
    \label{fig:c2_pcms}
\end{figure}

Three-dimensional calculations are rather time-consuming, and their numerical precision could also be problematic for implementation when investigating experimental data. Instead, we aimed to find an approximation that was precise and fast enough to be utilized in actual experimental analyses. Our approach here was that we fixed a set of parameters ($\alpha,R_1,R_2,R_3$) and evaluated the integral at $100^3$ points in momentum space. This gave us a fine enough resolution in momentum space for comparison purposes. First, let us compare the two-particle correlation functions in the PCMS. In Figure~\ref{fig:c2_pcms}, we can see the Bose--Einstein correlation functions with Coulomb interactions (full BEC) and without any final-state interactions (free BEC) from our 3D calculation and from the 1D calculation with quadratic and arithmetic average scale parameters and the angle averaged values of the 3D calculation. In the spherical case, on the left-hand plot, everything was as we would expect; however, on the right-hand plot, when we had a non-spherical source for the 3D calculation, we can see that there was a large difference between the correlation functions, both in the Coulomb interacting and in the free case.

\begin{figure}
    \includegraphics[width=0.5\textwidth]{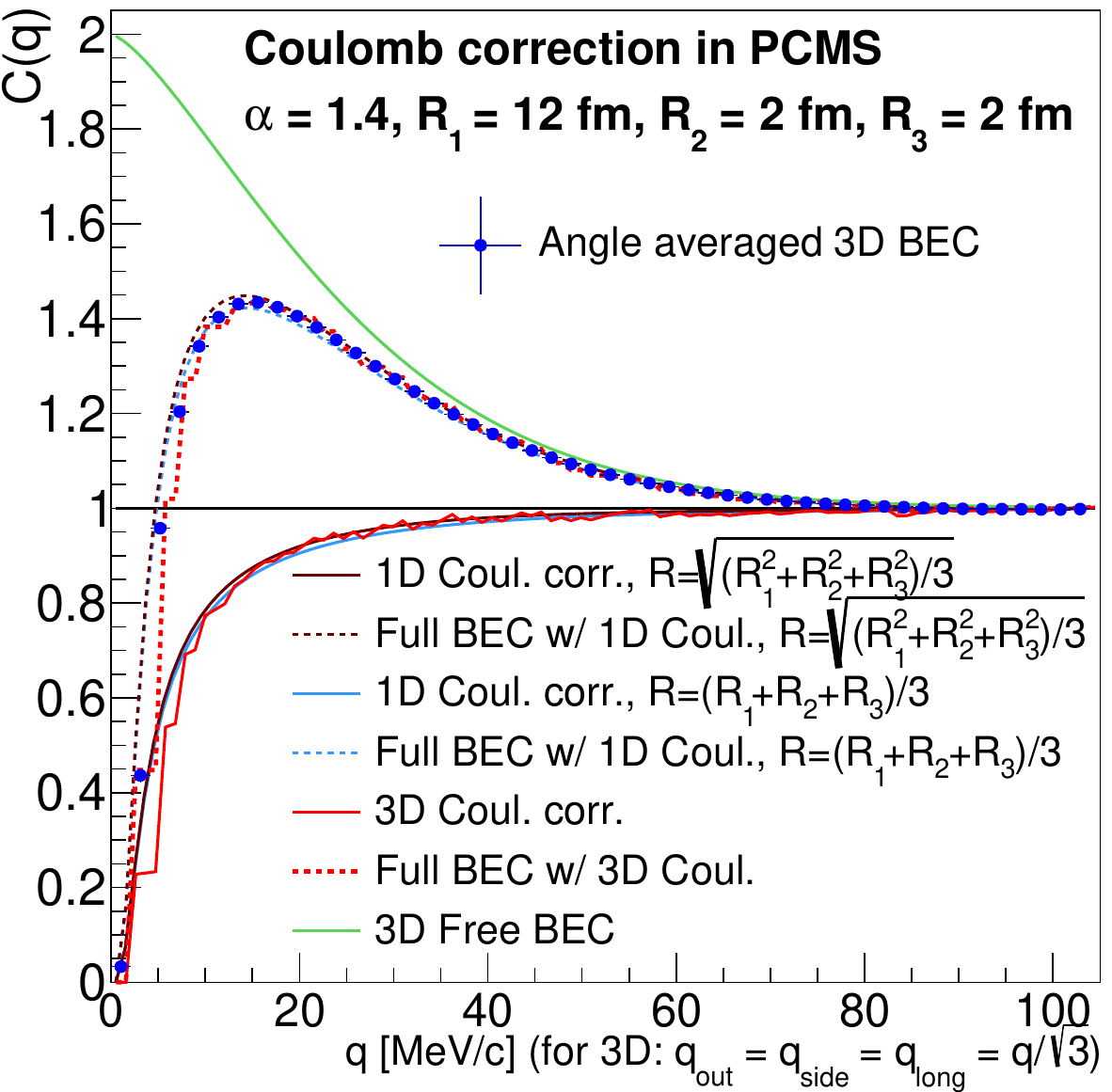}\hfill
    \includegraphics[width=0.5\textwidth]{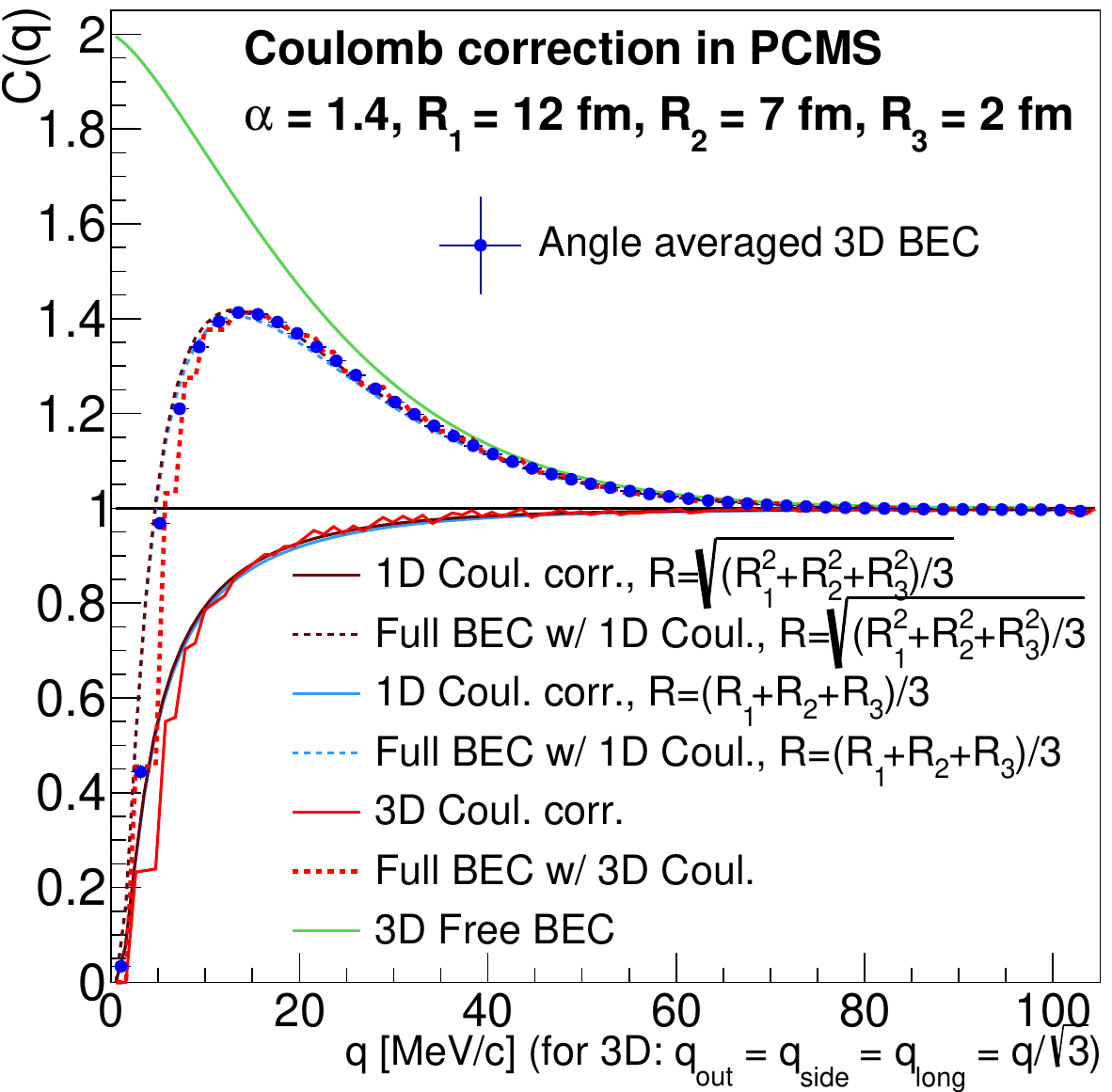}
    \caption{The Coulomb corrections and the Coulomb-corrected three-dimensional two-particle correlation function is shown in two non-spherical cases.}
    \label{fig:k_pcms}
\end{figure}

However, we were interested in the question of whether we could use the 1D calculation for the purposes of Coulomb correction only, viz., the ratio of the full and free BEC functions ($K=C_2^{(C)}/C_2^{(0)}$). One can see the comparison of Coulomb corrections in Figure~\ref{fig:k_pcms} with two sets of non-spherical parameters. The full BEC functions are here the Coulomb-corrected three-dimensional correlation functions (full BEC $=K\cdot C_{2,3D}^{(0)}$). The one-dimensional Coulomb corrections were evaluated at $|\vec{q}|$ in the PCMS, i.e., at $q_\text{inv}$ and at an average $R$ for $R_1,R_2$ and $R_3$. Although the correlation functions were quite different, we can see that the Coulomb corrections were very much the same. Now, we would like to point out the fact that one-dimensional and three-dimensional Coulomb corrections are very similar; therefore, in an experimental analysis, it is sufficient to use a one-dimensional Coulomb correction with the right parameter values. The error caused by the spherical Coulomb correction could be estimated, but it was not in the scope of this paper to give a quantitative limit for this uncertainty.

The application of the Coulomb correction in three-dimensional analyses is quite straightforward. If the measurement is in the LCMS and one has the momenta $q^L=q^L_\text{out},q^L_\text{side},q^L_\text{long}$ and the L\'evy scale parameters $R_\text{out},R_\text{side},R_\text{long}$ for particles with an average transverse momentum of $K_T$, which gives $\beta_T$, then one proceeds as follows. We used the assumption that the Coulomb correction transformed as a scalar. We evaluated the Coulomb correction (which was calculated in the PCMS) at momenta $q^P=(\sqrt{1-\beta_T^2}q^L_\text{out},q^L_\text{side},q^L_\text{long})$ and scale parameters $R_1=R_\text{out}/\sqrt{1-\beta_T^2}$, $R_2=R_\text{side}$, and $R_3=R_\text{long}$.

Accordingly, we used $q_\text{inv}=\sqrt{(1-\beta_T^2)q^{L2}_\text{out}+q_\text{side}^{L2}+q_\text{long}^{L2}}$ and an average of $R_1$, $R_2$, and $R_3$ when we used a 1D Coulomb correction. For example, we could use the quadratic average:
\begin{equation}\label{eq:r_avg_3d}
    R_{\rm PCMS}=\sqrt{\frac{1}{3}\left(\frac{R_\text{out}^2}{1-\beta_T^2}+R_\text{side}^2+R_\text{long}^2\right)}.
\end{equation} 

Therefore, the Coulomb correction that could be applied in a three-dimensional measurement was the following:
\begin{equation}
    K_\text{3D} = \frac{C_{2,1D}^{(C)}(q_\text{inv},R_\text{PCMS},\alpha)}{1+\exp{\left(-|q_\text{inv}R_\text{PCMS}|^{\alpha}\right)}},
\end{equation}
where $C_{2,1D}^{(C)}$ is the result from the integral of Equation~(\ref{eq:final_int}) in a spherical case with a radius of $R_\text{PCMS}$ according to Equation~(\ref{eq:r_avg_3d}) and at momentum $q_\text{inv}$, which can be calculated for every point in a three-dimensional measurement in the LCMS.

\subsection{Spherical (One-Dimensional) HBT Measurements}

Below, we investigate the implications of our calculations for one-dimensional HBT measurements. When we performed a one-dimensional measurement in the LCMS, we assumed that the source was spherical in this frame, i.e., $R=R_\text{out}=R_\text{side}=R_\text{long}$, and we had a single momentum variable $q_{\rm LCMS}=\sqrt{q^{L2}_\text{out}+q_\text{side}^{L2}+q_\text{long}^{L2}}$. But the Coulomb correction was calculated in the PCMS with $R_1,R_2,R_3$. This meant that a spherical source in the LCMS would imply a non-spherical ($R_1=R/\sqrt{1-\beta_T^2}$, $R_2=R_3=R$) source in the PCMS and the need for a three-dimensional Coulomb correction. However, we saw above that the non-spherical Coulomb correction could be well approximated with a spherical Coulomb correction if we used the right average R, viz., instead of ${R}_{\rm LCMS}=R$, we had to use

\begin{equation}\label{eq:r_pcms}
    {R}_{\rm PCMS}=\sqrt{\frac{1-\frac{2}{3}\beta_T^2}{1-\beta_T^2}}R,
\end{equation}
if we used a quadratic average R. Another problem stemmed from the fact that we could not reconstruct $q_\text{inv}$ from $q_{\rm LCMS}$. An obvious solution would be to measure all momentum variables instead of just the length of the momentum difference, but then the advantage of the 1D measurement over the 3D measurement (the possibility of a measurement with higher statistical significance) would be lost. We could try to overcome this obstacle in some other ways. One solid approximation could be the following: measure an $A(q_{\rm LCMS},q_\text{inv})$ distribution of particle pairs, and then use this to obtain a weighted Coulomb-correction, as shown below.
\begin{equation}\label{eq:weight_k}
K_\text{weighted}(q_{\rm LCMS}) = \frac{\int A(q_{\rm LCMS},q_\text{inv}) K(q_\text{inv}) 
\mathrm{d}q_\text{inv}}{\int A(q_{\rm LCMS},q_\text{inv}) \mathrm{d}q_\text{inv}}.
\end{equation}

\begin{figure}
    
    \includegraphics[width=0.5\textwidth]{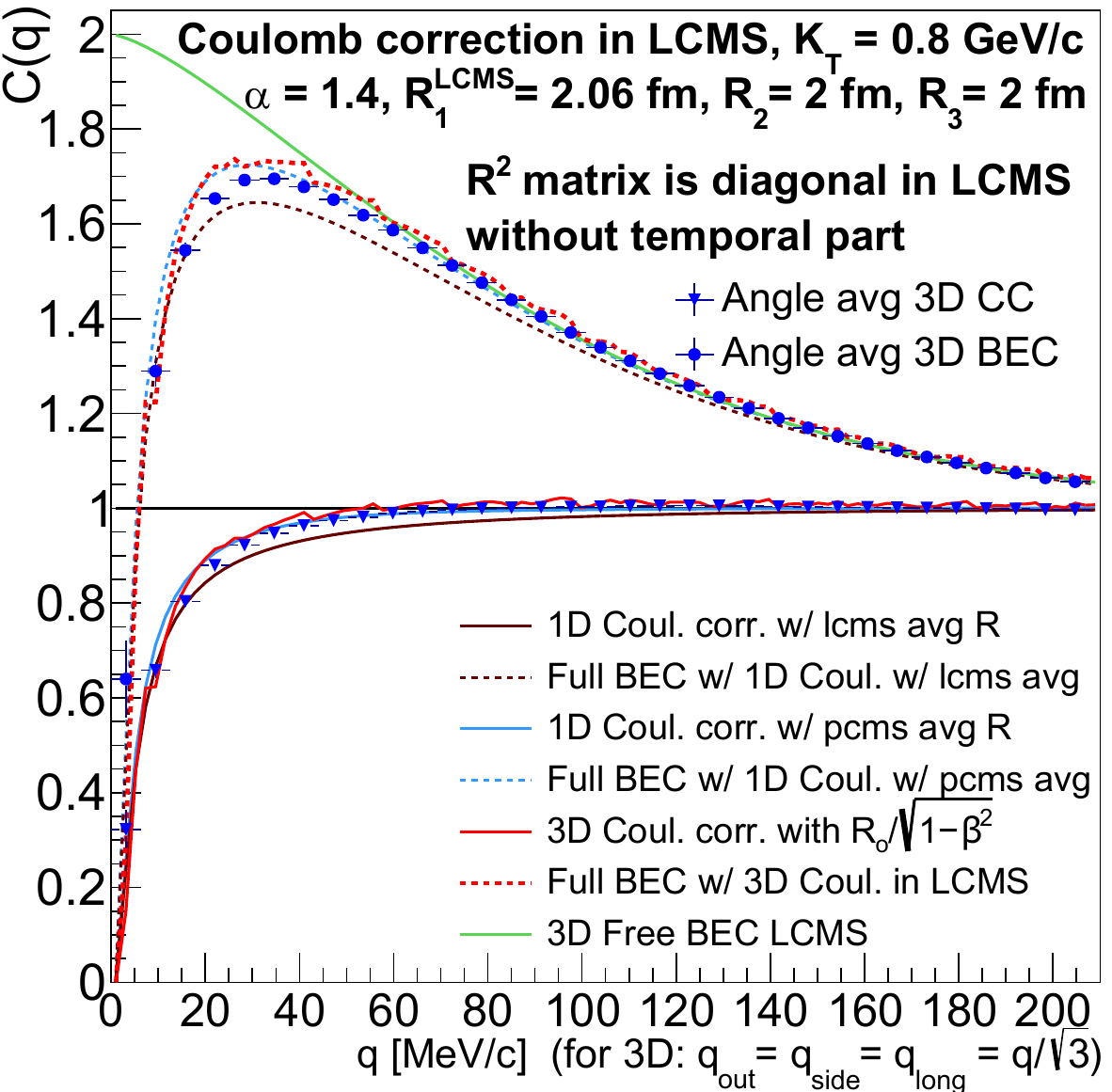}\hfill
    \includegraphics[width=0.5\textwidth]{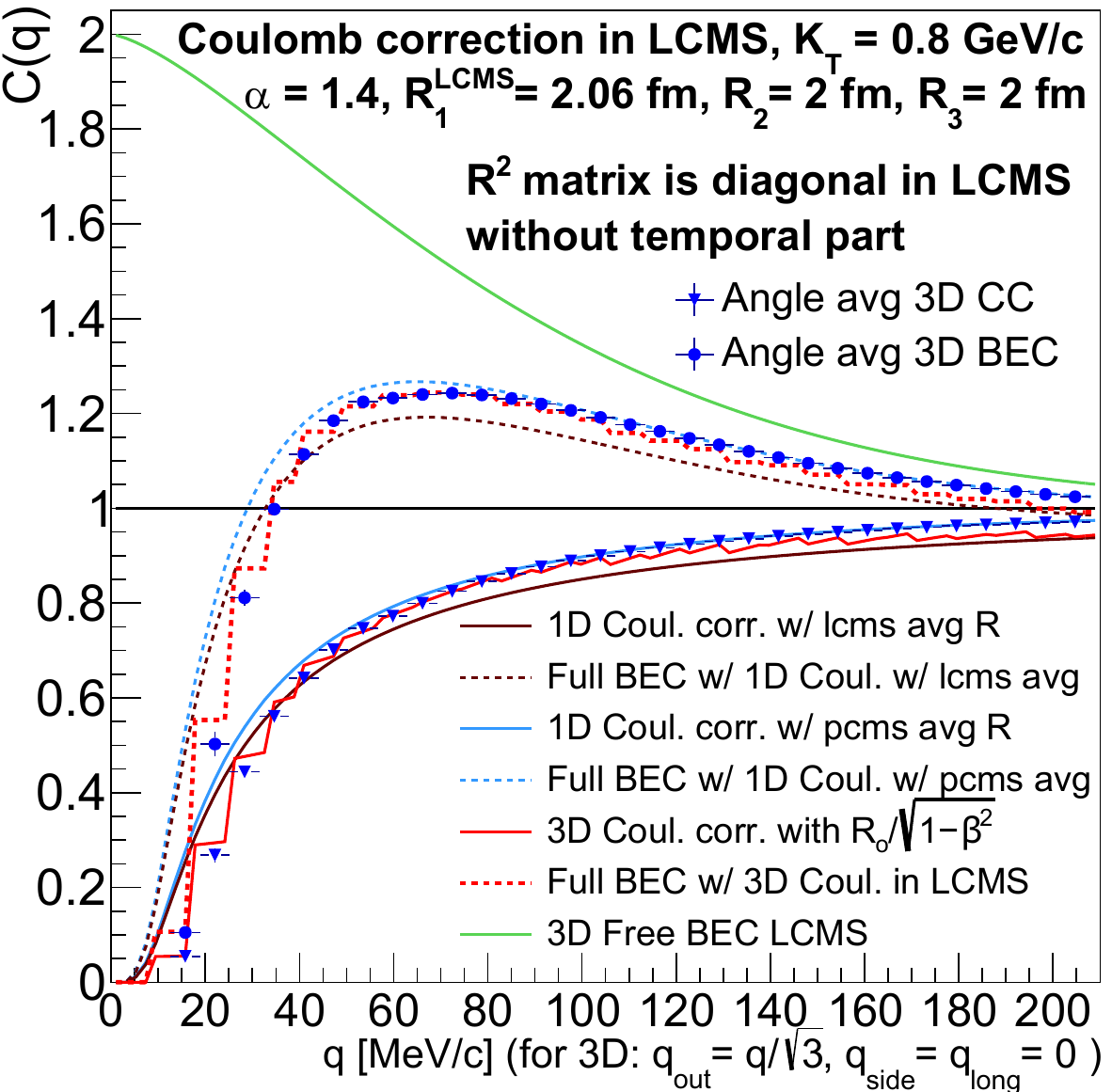}
    \caption{The Coulomb corrections and the Coulomb-corrected three-dimensional two-particle correlation function are shown in the LCMS when the source was spherical in the LCMS but 
    not for the calculation. On the left-hand side, we took the three-dimensional Coulomb correction along a diagonal line, and on the right-hand side along the $q_\text{out}$ 
    axis.}
    \label{fig:k_lcms}
\end{figure}

In Figure~\ref{fig:k_lcms}, we see the Coulomb correction and the corrected three-dimensional two-particle correlation functions for $K_T=0.8$ GeV/$c$ in the LCMS. The parameters were chosen so that in the LCMS we had an approximately spherically symmetric source ($R_\text{out}=2.06$ fm, $R_\text{side}=R_\text{long}=2$ fm). We can see that there was a clear difference between the two one-dimensional corrections, with one having an LCMS average R and the other having an average in accordance with Equation~\eqref{eq:r_pcms}. In the low-$q$ region, there was some difference between the angle-averaged, one-dimensional, and three-dimensional Coulomb corrections. Also, the numerical precision of the three-dimensional calculation made it challenging to decide between the options. However, we can see that from $q>20$ MeV/$c$, the angle-averaged and the three-dimensional Coulomb correction were in good agreement with the one-dimensional Coulomb correction with the average R of Equation~(\ref{eq:r_pcms}), and there was a consistent difference compared to the other one. The fact that the angle-averaged case was most similar to the one-dimensional case with the transformed average $R$ of Equation~\eqref{eq:r_pcms} indicated that using the latter for one-dimensional measurements was best. On the left-hand side, the three-dimensional correlation function was taken at a diagonal line in the LCMS ($q_\text{out}=q_\text{side}=q_\text{long}$), and on the right-hand side along the out axis. We did not rely on a weighted average for the one-dimensional Coulomb correction, as we could calculate $q_\text{inv}$.

Let us list the possible approaches to deal with Coulomb interactions in one-dimensional measurements carried out in an LCMS. We only list the options that make use of a one-dimensional calculation for the integral in Equation~(\ref{eq:final_int}); in these cases, the factor of Ref.~\cite{Csanad:2019lkp} can be used. A simpler solution would be to use the Gamow factor, where the source size is neglected. The most sophisticated approach would be to use the angle-averaged Coulomb correction from a three-dimensional calculation, but this would be an overly complex solution. The possibilities for making use of a one-dimensional Coulomb integral calculation are the following, ordered by increasing sophistication:

\begin{enumerate}
    \item Simply use $C_2^{(C)}(q_{\rm LCMS},R_{\rm LCMS})$, which means that one formally substitutes $q_{\rm LCMS}=q_\text{inv}$ and $R_{\rm PCMS}=R_{\rm LCMS}$.
    \item Take into account the fact that $q_{inv}\neq q_{\rm LCMS}$ but neglect the same for the scale parameters, and use the weighting method of Equation~(\ref{eq:weight_k});
    however, implement this not for the Coulomb correction, but for the correlation function instead. Thus, use the following formula in the fitting:
    
    \begin{equation}\label{eq:weight_c}
C_\text{2,weighted}(q_{\rm LCMS},R_{\rm LCMS})
= \frac{\int A(q_{\rm LCMS},q_\text{inv})C_2(q_\text{inv},R_{\rm LCMS})
  \mathrm{d}q_\text{inv}}
  {\int A(q_{\rm LCMS},q_\text{inv}) \mathrm{d}q_\text{inv}}.
    \end{equation}
    \item Following the same approach as above, use $R_{\rm LCMS}$ for the Coulomb correction and use a weighted average, though for the Coulomb correction this time. This approach is more sensible if one considers Figure~\ref{fig:c2_pcms}, where we saw that the correlation functions could look rather different even if in Figure~\ref{fig:k_pcms} the Coulomb corrections looked very much the same. Now, one uses $K_\text{weighted}(q_{\rm LCMS},R_{\rm LCMS})\cdot C^{(0)}_2(q_{\rm LCMS},R_{\rm LCMS})$ for fitting.\label{opt:3}
    \item One improvement to the methods mentioned above would be to consider the transformation of scale parameters; thus, use the average as in Equation~(\ref{eq:r_pcms}). The simpler version is the same as no.~\ref{opt:3}\ above, i.e., weighing the correlation function and using $C_{2,\text{weighted}}(q_{\rm LCMS},R_{\rm PCMS})$ for fitting. Here, however, one loses the explicit form of $C_2^{(0)}$ in the LCMS, which is known.
    \item The most sophisticated option would be to use $R_{\rm PCMS}$ only for the Coulomb correction and use the weighting of Eq.~(\ref{eq:weight_k}). The function used for fitting is now expressable as $K_\text{weighted}(q_{\rm LCMS},R_{\rm PCMS})\cdot C^{(0)}_2(q_{\rm LCMS},R_{\rm LCMS})$.
    \label{opt:best}
    \item Finally, an approach that is easier to implement than the previous methods making use of a distribution $A(q_{\rm LCMS},q_\text{inv})$ is to make an approximation for the $q_{\rm LCMS}$-$q_\text{inv}$ relationship that is appropriate for the Coulomb correction. One could be motivated by the left-hand plot of Figure~\ref{fig:k_lcms}, as the one-dimensional Coulomb correction with $R_{\rm PCMS}$ and the angle-averaged three-dimensional calculation were in relatively good agreement. The relationship $q_\text{inv}=\sqrt{1-{\beta_T^2}/{3}}q_{\rm LCMS}$ could be used, as it would hold for the diagonal line $q_\text{out}=q_\text{side}=q_\text{long}$. Therefore, the function that could be used for fitting would be $K(\sqrt{1-{\beta_T^2}/{3}}q_{\rm LCMS},R_{\rm PCMS})\cdot C^{(0)}_2(q_{\rm LCMS},R_{\rm LCMS})$.
\end{enumerate}

Additionally, either the distribution of particle pairs from same events (usually denoted with $A$) or some background distribution that has no quantum-statistical effects ($B$) could be used for weighting $C_2$ and $K$ \cite{Adare:2017vig}. Here, one could argue in favor of the latter; however, it is expected to make a small difference. The soundest approach for one-dimensional analyses is no.~\ref{opt:best} in the above list. 

\section{Conclusions}

We investigated Coulomb interactions for HBT measurements in the presence of L\'evy sources. Our results can be applied to three-dimensional and one-dimensional measurements alike. The results also hold for Gaussian or Cauchy sources, because these are special cases of the L\'evy source ($\alpha=2$ for Gaussian and $\alpha=1$ for Cauchy). We learned that a one-dimensional Coulomb correction could be reasonably effectively applied for three-dimensional measurements if we used the appropriately defined average of the three directional scale parameters (as in Equation~(\ref{eq:r_avg_3d}) above) and implemented the $q_\text{inv}$ invariant momentum difference as the momentum variable for the Coulomb correction. For one-dimensional measurements in the LCMS frame, we saw that one should use the average scale parameter as defined in Equation~(\ref{eq:r_pcms}) and evaluate the Coulomb correction at $q_\text{inv}$ as we calculated this in the PCMS frame, which in practice could be estimated with a weighted Coulomb correction according to option no.~\ref{opt:best} in the previous section. The above-detailed treatment of Coulomb interactions in heavy-ion collisions could be readily applied to experimental measurements. To our knowledge this indeed has now been achieved in several analyses from SPS through RHIC to LHC, based on the technique outlined in this paper~\cite{Lokos:2022exf,NA61SHINE:2023qzr,Mukherjee:2023hrz,Porfy:2023yii,Korodi:2023fug,CMS:2023xyd,Kovacs:2023temp}.

\section*{Acknowledgements}
The authors would like to express gratitude for the support of the Hungarian NKFIH grant No. K-138136.


\begin{thebibliography}{23}

\bibitem{Csorgo:1999sj}
T.~Cs{\"o}rg{\"o}, \emph{Particle Interferometry from 40 Mev to 40 TeV}
  (Springer, Dordrecht, 2000), Vol. 554 of \emph{NATO Science Series C:
  Mathematical and Physical Sciences}, chap.~8, pp. 203--257, ISBN
  978-94-011-4126-0, \texttt arXiv:{hep-ph/0001233}

\bibitem{Bolz:1992hc}
J.~Bolz, U.~Ornik, M.~Plumer, B.~Schlei, R.~Weiner, Phys.Rev. \textbf{D47},
  3860 (1993)

\bibitem{Adamczyk:2014mxp}
L.~Adamczyk et~al. (STAR Collaboration), Phys. Rev. \textbf{C92}, 014904 (2015), \texttt
  arXiv:{1403.4972}

\bibitem{Adare:2017vig}
A.~Adare et~al. (PHENIX Collaboration), Phys. Rev. \textbf{C97}, 064911 (2018),
  \texttt arXiv:{1709.05649}

\bibitem{Adler:2006as}
S.S. Adler et~al. (PHENIX Collaboration), Phys. Rev. Lett. \textbf{98}, 132301
  (2007), \texttt arXiv:{nucl-ex/0605032}

\bibitem{PHENIX:2007grx}
S.~Afanasiev et~al. (PHENIX Collaboration), Phys. Rev. Lett. \textbf{100}, 232301 (2008),
  \texttt arXiv:{0712.4372}

\bibitem{Shapoval:2013bga}
V.M. Shapoval, Y.M. Sinyukov, I.A. Karpenko, Phys. Rev. C \textbf{88}, 064904
  (2013), \texttt arXiv:{1308.6272}

\bibitem{NA61SHINE:2023qzr}
H.~Adhikary et~al. (NA61/SHINE Collaboration) (2023), \texttt arXiv:{2302.04593}

\bibitem{CMS:2023xyd}
A.~Tumasyan et~al. (CMS Collaboration) (2023), \texttt arXiv:{2306.11574}

\bibitem{Kurgyis:2018zck}
B.~Kurgyis (for the PHENIX Collaboration), Acta Phys. Polon. Supp. \textbf{12}, 477
  (2019), \texttt arXiv:{1809.09392}

\bibitem{Sinyukov:1998fc}
Y.~Sinyukov, R.~Lednicky, S.V. Akkelin, J.~Pluta, B.~Erazmus, Phys. Lett.
  \textbf{B432}, 248 (1998)

\bibitem{Bowler:1991vx}
M.G. Bowler, Phys. Lett. \textbf{B270}, 69 (1991)

\bibitem{Csanad:2019lkp}
M.~Csan\'ad, S.~L\"ok\"os, M.~Nagy, Phys. Part. Nucl. \textbf{51}, 238 (2020),
  \texttt arXiv:{1910.02231}

\bibitem{Csorgo:2003uv}
T.~Cs\"org\H{o}, S.~Hegyi, W.A. Zajc, Eur. Phys. J. \textbf{C36}, 67 (2004),
  \texttt arXiv:{nucl-th/0310042}

\bibitem{Pratt:1990zq}
S.~Pratt, T.~Cs\"org\H{o}, J.~Zim\'anyi, Phys.Rev. \textbf{C42}, 2646 (1990)

\bibitem{Bertsch:1988db}
G.~Bertsch, M.~Gong, M.~Tohyama, Phys. Rev. \textbf{C37}, 1896 (1988)

\bibitem{Landau3:1977}
L.D. Landau, L.M. Lifshitz, \emph{Quantum Mechanics Non-Relativistic Theory,
  Third Edition: Volume 3}, 3rd~edn. (Pergamon, 1977), ISBN 978-0-08-020940-1

\bibitem{Metropolis:1953am}
N.~Metropolis, A.W. Rosenbluth, M.N. Rosenbluth, A.H. Teller, E.~Teller, J.
  Chem. Phys. \textbf{21}, 1087 (1953)

\bibitem{Hastings:1970aa}
W.K. Hastings, Biometrika \textbf{57}, 97 (1970)

\bibitem{Lokos:2022exf}
S.~L\"ok\"os, Acta Phys. Polon. Supp. \textbf{15}, 30 (2022), \texttt
  arXiv:{2206.13952}

\bibitem{Mukherjee:2023hrz}
A.~Mukherjee, Universe \textbf{9}, 300 (2023), \texttt arXiv:{2306.13668}

\bibitem{Porfy:2023yii}
B.~Porfy (for the NA61/SHINE Collaboration), Universe \textbf{9}, 298 (2023), \texttt arXiv:{2306.08696}

\bibitem{Korodi:2023fug}
B.~K\'orodi (for the CMS Collaboration), Universe \textbf{9}, 318 (2023), \texttt arXiv:{2306.16353}

\bibitem{Kovacs:2023temp}
L.~Kov\'acs (for the PHENIX Collaboration), accepted at Universe (2023) \textbf{9} (2023)

\end{thebibliography}
\end{document}